\documentclass[12pt]{article}
\usepackage{graphicx}

\title{Quantitative Analysis of Genealogy Using Digitised \mbox{Family} Trees}
\author{Michael Fire$^{1}$, Thomas Chesney$^2$ \& Yuval Elovici$^1$}

\begin{document}
\maketitle

\begin{enumerate}
\item Telekom Innovation Laboratories at Ben-Gurion University
\item Nottingham University Business School
\end{enumerate}

\begin{abstract}
Driven by the popularity of television shows such as Who Do You Think You Are? many millions of users have uploaded their family tree to web projects such as WikiTree \cite{Wikitree14}. Analysis of this corpus enables us to investigate genealogy computationally. The study of heritage in the social sciences has led to an increased understanding of ancestry and descent \cite{Zerubavel12} but such efforts are hampered by difficult to access data \cite{Duff03}. Genealogical research is typically a tedious process involving trawling through sources such as birth and death certificates, wills, letters and land deeds \cite{Cortada}. Decades of research have developed and examined hypotheses on population sex ratios, marriage trends, fertility, lifespan, and the frequency of twins and triplets. These can now be tested on vast datasets containing many billions of entries using machine learning tools. Here we survey the use of genealogy data mining using family trees dating back centuries and featuring profiles on nearly 7 million individuals based in over 160 countries. These data are not typically created by trained genealogists and so we verify them with reference to third party censuses. We present results on a range of aspects of population dynamics. Our approach extends the boundaries of genealogy inquiry to precise measurement of underlying human phenomena.
\end{abstract}

Genealogy is the study of family origins, bloodlines, and history. Recent advances in web technologies have encouraged millions of amateur genealogists to discover, assemble and share their family history by constructing their own online family trees. These joint effort online genealogy websites create large-scale data containing billions of entries regarding people who lived in past centuries \cite{Geni14, Fire14} and there is a lot of interest in contemporary relationships too, as evidenced by the Global Family Reunion campaign \cite{GFR14}. The resulting corpora consist of personal information on family members going back many generations, and they provide details for each individual such as their date and place of birth and death, and their place in their family tree. These data create unique opportunities to study many and various aspects of human life and of the human life cycle over the past centuries.

The data we use were provided by WikiTree, a free, collaborative worldwide family tree project created by a community of amateur genealogists. Data are available on 6.67 million people in over 160 countries (but mainly the US, UK, Germany, Canada, New Zealand and Holland) going as far back as the first century. We coded relationships between individuals as either spouse, child, parent or sibling, and where available additional personal data were attached recording sex, the year and country of birth and death, and marriage date and location. The data therefore have three main dimensions: time, location and personal characteristics. Limits were set on personal characteristics and values falling outside of them (such as an age at death of 122+) were replaced with a missing value placeholder. Data were validated by WikiTree using their in-house procedures which include checking source materials and by making individuals' profiles editable only by a limited list of users, and we provided additional validation by comparing lifespans in the data with those reported by third party sources \cite{ONS10, Mitchell01}.

These data allow for many analyses that would be cumbersome or impossible using traditional genealogy research. For example one traditional study \cite{Gogele11} examined the lifespans of 53,000 people, reconstructing pedigrees on 1,000 living Italians back to the early 17th century using family books and parish registers back to 1924 and municipality lists thereafter. The parish registers, which record baptisms, marriages, and burials, were viewed on microfilm and researchers had to physically attend the archive at Bolzano. Assembling these data took two years. We were able to do a similar study in under a week.

\textbf{Name trends.} To start we can illustrate the use of the data by highlighting trends in given names alongside cultural events. Figure 1 plots the ratio of the number of times selected given names were used in a year divided by the total number of babies born in a year against time. The graph for `Wendy' for instance grows with the popularity of the Peter Pan story -- originally staged in 1904, with the novel published in 1911 and related films and books appearing from the 1920s onwards up to Leonard Bernstein's 1950 musical and the Disney film in 1953. Contrary to popular belief the name Wendy existed prior to Peter Pan.

%\textbf{TC to redo the figure and caption.}

Another interesting trend is variation in the most common child's name. From the year 1000 up to 2000 the ratio of unique given names used per decade (where each name appears at least 10 times in each decade--other names were ignored) is shown in Figure 2. The chart shows low variation during the High Middle Ages and the Victorian era, suggesting that the desire to pick an uncommon name for children \cite{Wattenberg05} is not new. The trend of naming a son after its father rises then falls through the 16th century, and throughout history there have been fewer girls named for their mother than boys named for their father (Figure 3). About 24\% of twins's names start with the same letter. The most frequent twin names between 1800 and 1900 are Mary and Martha, and John and James. 

\textbf{Births and fertility.} We plotted the average age at which mothers gave birth to their first and last born against time in decade sized bins (Figure 4). The graphs show an upward trend in both as women tended to wait longer before having children. We then examined the frequency of twins and triplets. Hellin's Law states that one in every 89 human maternities is twins and one in every $89^2$ is triplets although a proof exists \cite{Fellman93} that this cannot hold as a general rule and many exceptions have been found \cite{Fellman09}. Nevertheless we assessed Hellin's Law with the WikiTree data using all births which occurred between 1800 and 1900 (where the bulk of recorded births occur) and found support that it at least approximates reality. Of 963,416 births, 10,246 were twins (0.0106\%), and 128 (0.00013\%). Twin gender ratios were almost even. Where gender data are available the sets of twins were: male-male -- 3,257 (32.7\%); female-female -- 3,376 (33.9\%); and male-female -- 3,307 (33.3\%).

The relationship between natural factors and human sex ratio remains an active area of scientific research and these data may be able to contribute to such efforts. Examining the gender ratios in our data we observe a small but steady rise in females from the middle ages on (Figure 5). We acknowledge the possibly---as others have---that this could be because births of men tended to be recorded more than those of females however in developed countries studies have found that the human sex ratio at birth has historically varied for natural reasons \cite{James08}.

\textbf{Marriages.} The age at which individuals first got married is shown in Figure 6. The general trend is in any given time period, for males to marry later than females, and the age increases over time. The raw data collaborates that during the medieval period it was not unknown for girls aged 12 and boys aged 14 to marry \cite{Decameron14} but the trend shown in the graph is that these young ages did not represent the average. 

\textbf{Lifespan.} Previous studies have found that spouses have an impact on an individual's lifespan \cite{Lillard95}. We find support for this -- if an individual's spouse lives longer, then that individual lives longer too: the age at death in years of one partner correlates with the age of death of the other with a Pearson coefficient of $r=0.224$. Twins also tend to have the same lifespan: the age at death of Twin 1 correlates with the age at death of Twin 2 with $r=0.22$.

\textbf{Computational genealogy.} Computational genealogy is the application of machine learning tools, graph analysis and related techniques to the analysis of high volume ancestry data and is an emerging branch of computational social science \cite{Lazer11}. The results are a new type of evidence in social science. Here we have given a brief survey of early findings but the field opens up many more possibilities including:

\begin{itemize}
\item Highlighting immigration trends, for instance by looking at surname changes between Italian and Irish immigration to the US.

\item Charting lifespan alongside economic trends -- genealogical data can highlight wars, disease outbreaks and vaccinations on the charts and may be able to quantify the impact and spread of vaccinations and other health innovations. We have found for example a peak in births around 9 months after the end of the American Civil war, and a peak in deaths at the time of the Battle of Flodden.

\item Assessing the impact of key religious events such as The Reformation (1517) on the frequency of marriages and births. 

\item Analysing how families moved and split geographically -- in the past did entire families (grown siblings) re-locate together? Was it common for children or grandchildren to move back to an ancestral family home?

\item Identifying genetic diseases by searching for patterns on age of death within the same family.

\item Combining two or more sources on a famous bloodline, for instance by using the WikiTree data alongside Wikipedia.
\end{itemize}

The results of such investigations will be of interest to a range of branches of social science.

\clearpage

\begin{figure}
\begin{centering}
\scalebox{0.4}{\includegraphics{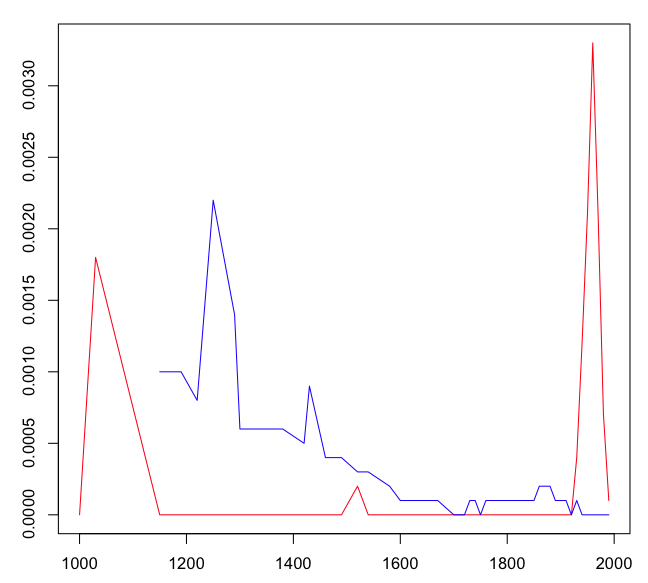}}
\caption{Analysis of the frequency with which given names appear in the dataset. The name Wendy is shown here in Red.  In blue is the name Adolf. The $y$ axis shows the ratio between all people with the given first name born in a decade divided by the total number of births with valid first name that were born in that decade.}
\end{centering}
\end{figure}

\begin{figure}
\begin{centering}
\scalebox{0.4}{\includegraphics{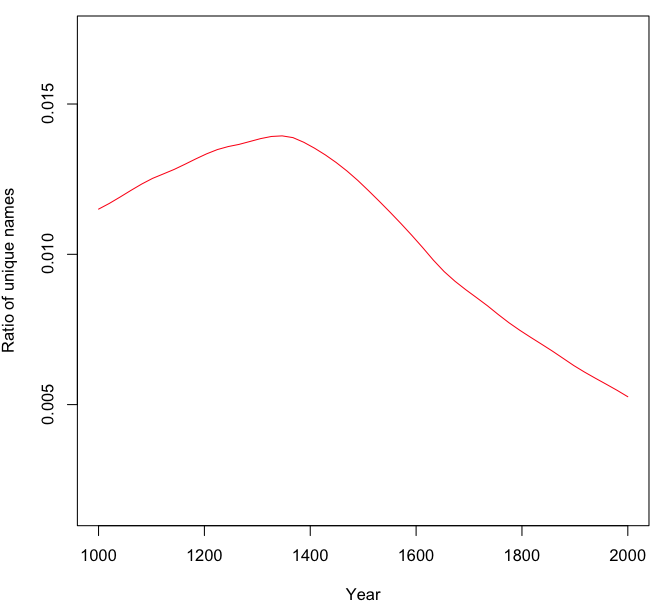}}
\caption{The ratio of unique given names used per decade (where each name appears at least 10 times in each decade)}
\end{centering}
\end{figure}

\begin{figure}
\begin{centering}
\scalebox{0.4}{\includegraphics{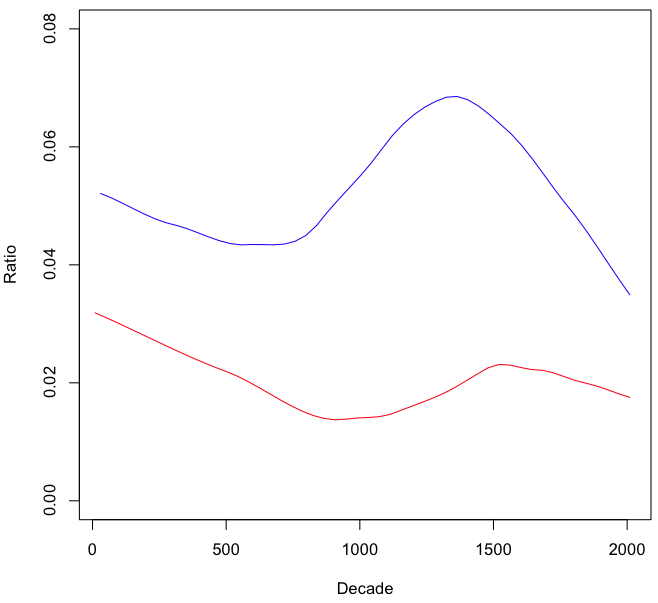}}
\caption{The blue line shows the ratio of sons named after their fathers, the red line shows the ratio of daughters named after their mothers.}
\end{centering}
\end{figure}

\begin{figure}
\begin{centering}
\scalebox{0.4}{\includegraphics{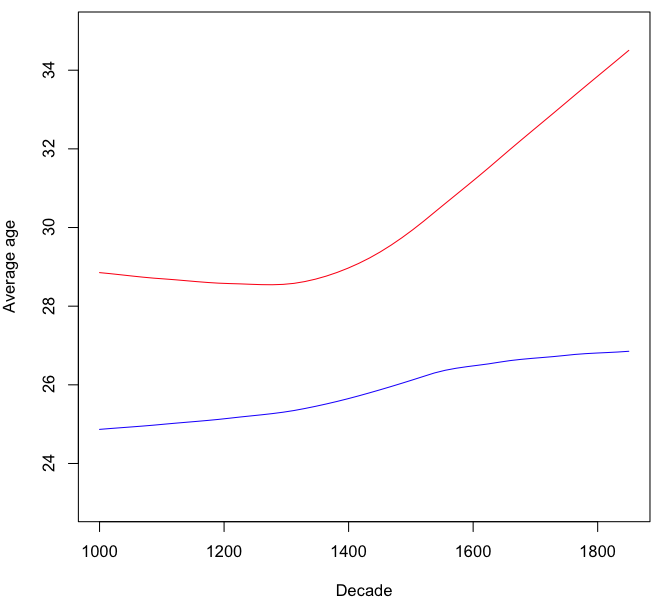}}
\caption{Mother's age at the time of her first born (blue) and last born (red).}
\end{centering}
\end{figure}

\begin{figure}
\begin{centering}
\scalebox{0.4}{\includegraphics{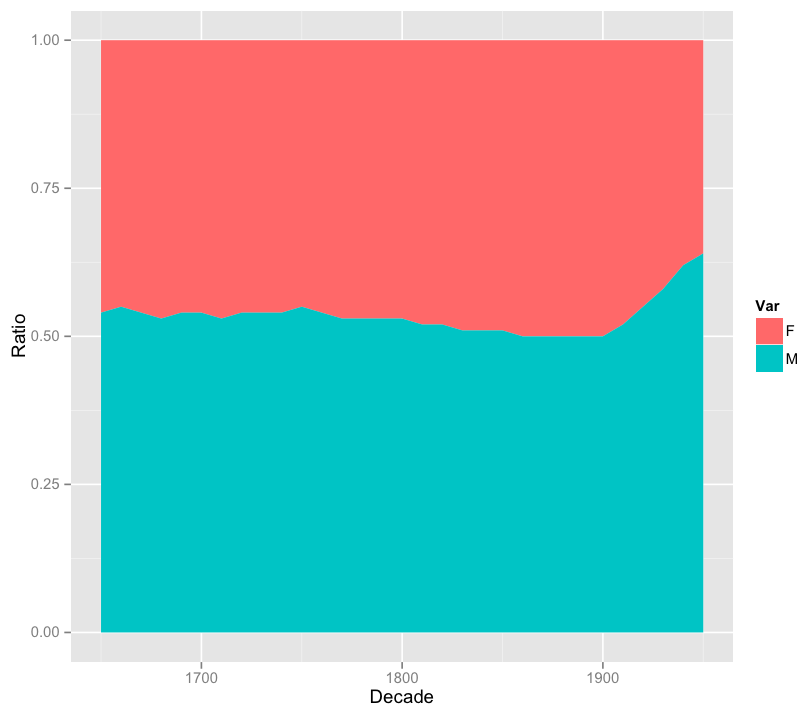}}
\caption{Genders ratio for births between 1650 and 1950.}
\end{centering}
\end{figure}

\begin{figure}
\begin{centering}
\scalebox{0.4}{\includegraphics{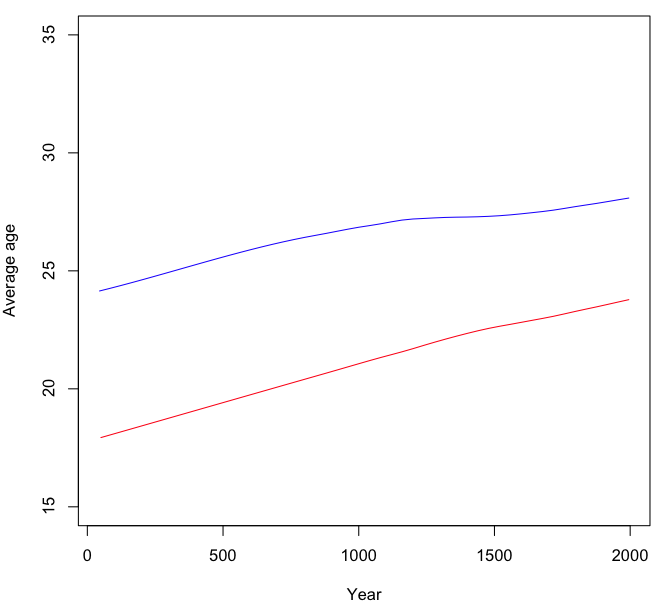}}
\caption{Average age that males (blue) and females (red) first got married.}
\end{centering}
\end{figure}

\clearpage

\bibliographystyle{naturemag}
\bibliography{Chesney}
\end{document}